# Influence of composition and heating schedules on compatibility of FeCrAl alloys with high-temperature steam

*Chongchong Tang [a,\*], Adrian Jianu [b], Martin Steinbrueck [a], Mirco Grosse [a], Alfons Weisenburger [b], Hans Juergen Seifert [a]*

[a]*Institute for Applied Materials (IAM), Karlsruhe Institute of Technology (KIT), D-76021 Karlsruhe, Germany*

[b]*Institute for Pulsed Power and Microwave Technology, Karlsruhe Institute of Technology(KIT), D-76021 Karlsruhe, Germany*

*Email: Chongchong.tang@kit.edu*

**Abstract:** FeCrAl alloys are proposed and being intensively investigated as alternative accident tolerant fuel (ATF) cladding for nuclear fission application. Herein, the influence of major alloy elements (Cr and Al), reactive element effect and heating schedules on the oxidation behavior of FeCrAl alloys in steam up to 1500°C was examined. In case of transient ramp tests, catastrophic oxidation, i.e. rapid and complete consumption of the alloy, occurred during temperature ramp up to above 1200°C for specific alloys. The maximum compatible temperature of FeCrAl alloys in steam increases with raising Cr and Al content, decreasing heating rates during ramp period and doping of yttrium. Isothermal oxidation resulted in catastrophic oxidation at 1400°C for all examined alloys. However, formation of a protective alumina scale at 1500°C was ascertained despite partial melting. The occurrence of catastrophic oxidation seems to be controlled by dynamic competitive mechanisms between mass transfer of Al from the substrate and transport of oxidizing gas through the scale both toward the metal/oxide scale interface.

## 1. Introduction

Zirconium-based alloys possess low neutron absorption cross section, good corrosion and irradiation resistance as well as high mechanical strength. These outstanding properties guarantee them being well qualified for utilization as state-of-the-art cladding and structural components in water-cooled nuclear reactors with respect to normal operation [1]. However, an undesirable limitation is their fast self-catalytic exothermic reaction with high-temperature steam in case of off-normal conditions. Once the environment inside the core changes from normal operating conditions to accident scenarios, e.g. loss of coolant accidents (LOCA), the Zr-based claddings first suffer ballooning and then bursting at around 700 - 900°C [2]. With continuous increase of the core temperature, the claddings undergo severe degradation and accelerated oxidation. Due to the exothermic zirconium-steam oxidation reaction, a large quantity of heat is released and hydrogen gas is produced increasing the risks of hydrogen detonation and radioactive fission products release [3–5]. Therefore, the development of advanced accident-tolerant fuels (ATFs) with larger safety margins became one primary focus after the Fukushima accident in 2011 [6,7]. Alternative ATF cladding concepts, including coated Zr-based cladding, hybrid ceramic/metal cladding, or advanced ceramic and metallic cladding, owning excellent high-temperature oxidation resistance are being investigated aiming to enhance the accident tolerance [2,6,8–11].

FeCrAl-based alloys represent one promising candidate as alternative ATF cladding material. These alloys show attractive properties, like good formability, high mechanical properties and, more specifically, excellent high-temperature oxidation resistance [12]. Traditional FeCrAl alloys with relative high chromium addition (~20 wt.%) and moderate Al concentration (~5 wt.%) are specifically optimized for high-temperature oxidizing applications, e.g. heating elements and catalytic converters [13]. The excellent oxidation resistance of FeCrAl-based alloys in oxidizing atmospheres relies on the growth of an external, dense and adherent $\alpha$-$Al_2O_3$ scale due to the selective oxidation of Al at elevated temperatures. The addition of around 20 wt.% Cr (known as third element effect, TEE) reduces the amount of Al needed to establish the protective alumina scale [14]. However, high Cr content in these alloys may trigger potential irradiation embrittlement at typical LWR operating temperatures during service [15,16]. In order to optimize the chemical composition of the FeCrAl-based alloys to make them suitable for nuclear fission application while maintaining their excellent oxidation resistance, the straightforward solution consisting in increasing the Al content and simultaneously decreasing the Cr content is not acceptable due to the pronounced Al negative effect on the workability and mechanical properties (especially ductility) [17]. Current studies are focused on preserving a delicate balance between Cr and Al content in conjunction with minor additions of other elements that eliminate or manipulate undesirable attributes [15,18–20].

Intensive studies have been performed to investigate the oxidation behavior of FeCrAl-based alloys in both dry and moist atmospheres. However, previous studies on investigating their oxidation behavior and mechanism were focused on steady-state condition at certain temperatures (i.e. after successful establishment of a protective $Al_2O_3$ scale), like breakaway oxidation, reactive element (RE) effect and stress development [21–24]. In addition, the examined temperatures were generally low (< 1300°C) considering nuclear application with respect to accident scenarios. A plenty of studies have been dedicated to examining their behavior in high-temperature steam for nuclear application recently [25–28]. The comprehensive understanding of underlying mechanisms on growth and failure of protective alumina scale, particularly during the initial stage of oxidation, at elevated temperatures is not yet satisfying. The purpose of this study was to elucidate the influence of both, the chemical composition, including major alloying elements and reactive element, and the heating schedules, on the compatibility of FeCrAl alloys with high-temperature steam. The final aim was to shed more insights into the response of this alloy system during postulated nuclear accident scenarios.

## 2. Experiment arrangement

Fourteen model FeCrAl-based alloys without or with yttrium (Y) addition were prepared as ingots from high purity element powders by arc melting under argon atmosphere. The small addition of Y (0.3 wt.%) in some alloys is expected to be in a solid solution state. These model alloys were in as-cast condition without additional heat treatment. In addition, two commercial alloys, Kanthal APM and D as tubes, were purchased from Kanthal/Sandvik, Sweden. The alloys nominal compositions together with the actual compositions determined by energy dispersive X-ray spectroscopy (EDS) are listed in Table 1. For oxidation tests, cylindrical disc specimens with around 1.2 mm thickness and ring specimens with around 4 mm length were cut from the ingots and the tubes, respectively. The diameter of the disc specimens was around 10 mm. The geometry of the two commercial alloy specimens is as follows: the outer diameter APM: 25 mm and D: 6.0 mm, the wall thickness APM: 1.6 mm and D: 0.4 mm. The Kanthal APM is a powder metallurgical product and both commercial alloys contain small amounts of additional alloying elements (Si, Mn and C). Prior to oxidation, the cutting ends of the specimens were deburred and ground using metallographic grinding with abrasive SiC paper down to 4200 grit. The specimens were then cleaned in ultrasonic solution following a sequence of water, acetone and isopropanol, finally dried in hot air.

The high-temperature steam oxidation experiments in this study were performed using a horizontal corundum tube furnace ((inner diameter: 32 mm, length: 600 mm) under normal pressure. The specimens were placed on a corundum crucible sample holder located in the center of the furnace. Two different kinds of tests were conducted: transient tests and isothermal tests. During both tests, the specimens were oxidized in a flowing gas mixture consisting of argon and steam. The gas composition and flow rates were defined by flow controllers and a CEM (Controlled Evaporation and Mixing) system both delivered by Bronkhorst®. The composition of the off-gas was *in-situ* analyzed by a quadrupole mass spectrometer (Balzers GAM300). Particular interest was focused on the evolution of hydrogen release rate in the off-gas during the tests, which was used as measurement for the oxidation rate. Pure steam atmosphere cannot be achieved due to argon is needed to serve as the carrier gas and reference gas for mass spectrometer analysis.

Table 1 Designation and actual composition of model FeCrAl-based alloys and commercial alloys investigated in this study (*values from Kanthal, Sandvik AB, Sweden)

| Type | Sample | Designation composition | Cr (wt.%) | Al (wt.%) | Other (wt.%) | Fe (wt.%) |
|---|---|---|---|---|---|---|
| Model Alloys | P1 | Fe6Cr6Al | 6.2 | 6.8 | - | Bal. |
| | P2 | Fe8Cr6Al | 8.4 | 6.6 | - | Bal. |
| | P3 | Fe10Cr5Al | 10.8 | 5.6 | - | Bal. |
| | P4 | Fe14Cr4Al | 14.5 | 4.2 | - | Bal. |
| | P5 | Fe16Cr4Al | 16.4 | 4.7 | - | Bal. |
| | P6 | Fe6Cr8Al | 6.4 | 8.8 | - | Bal. |
| | P7 | Fe10Cr7Al | 10.4 | 7.6 | - | Bal. |
| | P8 | Fe12Cr7Al | 12.4 | 7.5 | - | Bal. |
| | P9 | Fe16Cr6Al | 16.9 | 6.4 | - | Bal. |
| | P10 | Fe12Cr5Al | 12.3 | 5.8 | - | Bal. |
| | P11 | Fe12Cr8Al | 12.2 | 8.2 | - | Bal. |
| | P12 | Fe16Cr8Al | 16.5 | 8.7 | - | Bal. |
| | P13 | Fe12Cr5Al0.3Y | 12.0 | 5.1 | ~0.31 Y | Bal. |
| | P14 | Fe16Cr8Al0.3Y | 16.4 | 8.5 | ~0.32 Y | Bal. |
| Commercial Alloys | K-APM* | Fe20Cr5.8Al | 20.5 | 5.8 | Minor Si, Mn, C | Bal. |
| | K-D* | Fe20Cr4.8Al | 20.5 | 4.8 | | Bal. |

Typical examples of temperature profile and gas concentrations during transient tests and isothermal tests are shown in Fig. 1. In the case of transient tests, the specimens were firstly heated in high purity Ar atmosphere (99.9999%) with a flow rate of 40 l/h from ambient temperature to 500°C, and maintained at 500°C for certain time to stabilize the temperature. Subsequently, the steam was introduced into the furnace by changing the gas flow to 20 l/h Ar and 20 g/h $H_2O$. The concentration of the steam during the oxidation phase was ~55 vol.%. The specimens were oxidized in steam whereas a transient ramp period heating to the desired temperatures (1450°C or 1300°C) followed by an isothermal phase with different dwell time. During the temperature ramp up, two different heating rates were adopted: 5 and 10 K/min. The temperature ramp rates on the progression of a specific accident scenario can vary significantly, from few to more than one hundred Kelvin per minute, depending on the accident severity [29]. The selection of these two heating rates took into account the actual accident conditions as well as the facility capabilities. After the isothermal exposure, the specimens were finally cooled down to room temperature in the furnace in Ar. With respect to the isothermal tests, the specimens were heated in Ar to the desired temperatures (1200-1500°C) with a fixed heating rate of 10 K/min, and subsequently isothermally oxidized in argon and steam mixture atmosphere for 1 hour or 15 min. The other parameters (gas flow rates and cooling rates) remained the same as in the transient test. The flow rates of gas shown herein are in standard conditions (25°C, 1bar).

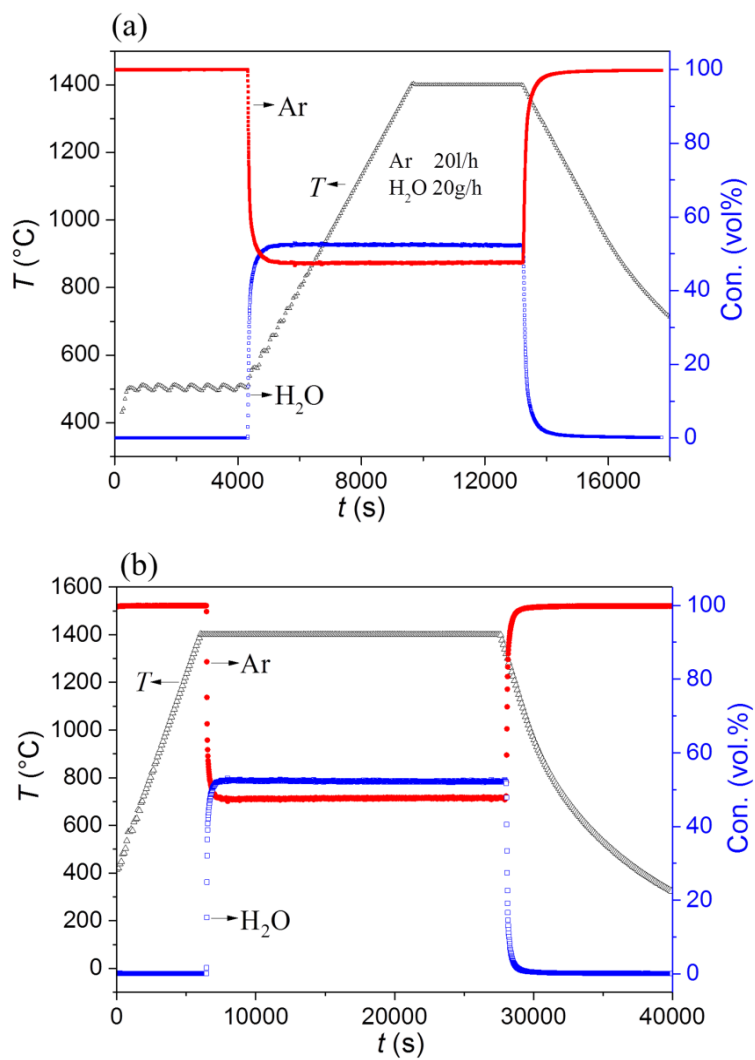

Fig. 1 Typical examples of the temperature profile and gas concentrations during (a) the transient tests and (b) the isothermal tests.

The mass of the specimens before and after oxidation was measured using an analytical balance with resolution of ±0.0001 g. The phase compositions of the surface after oxidation were characterized using a step-scanning X-ray diffraction (XRD, Seifert PAD II) with Cu Kα1 radiation ($\lambda$=0.154 nm) in conventional $\theta$-$2\theta$ geometry. The samples which were completely oxidized were ground to powder for XRD measurements. The surface and cross section were examined using a field-emission scanning electron microscope (SEM, Philips XL30S), equipped with an energy-dispersive X-ray spectroscopy (EDS) for elemental analysis.

# 3. Results

## 3.1 Transient tests with 5 and 10 K/min heating rates

### 3.1.1 Experiments up to 1450°C

Fig. 2 shows the hydrogen production rates for several representative alloys during the transient tests from 500 to 1450°C with two different heating rates and subsequent holding for 1 h at 1450°C. Based on the hydrogen production behavior recorded by mass spectrometer during the tests and the surface morphology after the tests (shown later), the samples can be classified into two categories: samples undergoing catastrophic oxidation and samples forming a protective oxide scale. The catastrophic oxidation phenomenon indicated that the samples failed and completely oxidized without surviving protective oxide scales. In contrast, the second category, where the samples survived and formed robust protective oxide scales, was not failed during the test.

The hydrogen curves demonstrate obviously different characteristics for the failed samples (catastrophic oxidation) and the survived samples (protective scale) during the transient tests, as shown in Fig.2. For the failed samples (as examples: P4 and P8 in Fig.2a, P10 in Fig.2b), the hydrogen concentration first maintained at a low level, then increased sharply to an extremely high value, more than two magnitudes higher, accompanied by a strong peak once the temperature reaching certain values (above 1200°C). Subsequently, it decreased quickly to the initial level measured during the heating up period, indicating that rapid and complete oxidation occurred. The final mass gains of this category alloys were around 170 mg/cm$^2$ corresponding to 47 wt.%. In case of the survived samples (P9 in Fig.2a, P9 and P13 in Fig.2b), the hydrogen production rates increased gradually during the heating up period and then decreased slightly in the isothermal oxidation period. However, the hydrogen production rates maintained at an extremely low level during the whole exposure as observed in the inserted images. The mass gains per unit area for those samples after the transient test were in the range of 2-4 mg/cm$^2$. These behaviors indicate that a protective oxide scale formed on the surface to inhibit direct contact of the external atmosphere with the alloy.

Table 2 presents the steam compatibility of all alloys investigated, as a function of composition and heating rate in the transient tests, based on the hydrogen evolution. The maximum tolerable temperature in steam, above which the oxide scales are supposed to fail, is defined as the temperature at which the sharp peak of the released hydrogen starts to emerge (Fig.2). The maximum tolerable temperature of the survived samples was marked as 1450°C (the highest temperature during the test) as shown in Table 2. The first category, i.e. samples suffered catastrophic oxidation, mainly comprised of model alloys with relatively low Cr and Al contents and without reactive element (Y), as well as the two commercial alloys tested at higher heating rate (10 K/min). Increasing the aluminum and chromium contents in the alloys, the maximum tolerable temperature to steam increased. By applying a lower heating rate during the ramp period, the maximum tolerable temperature also raised (Table 2). One meaningful finding is that the reactive element (Y) could significantly improve the high-temperature steam tolerance of the FeCrAl-based alloys during the transient test. The P10 sample, with about 12 wt.% Cr and 5 wt.% Al, failed at 1330°C with 10 K/min heating rate. However, with additional 0.3 wt.% Y (P13), the alloy could withstand temperature up to 1450°C and survived during the test. An obvious reduction of the Cr and Al contents can be achieved by a small amount of Y dopant. The two commercial alloys with relatively high Cr content succeeded to 1450°C with 5 K/min heating rate, but failed below 1400°C with a higher heating rate (10 K/min).

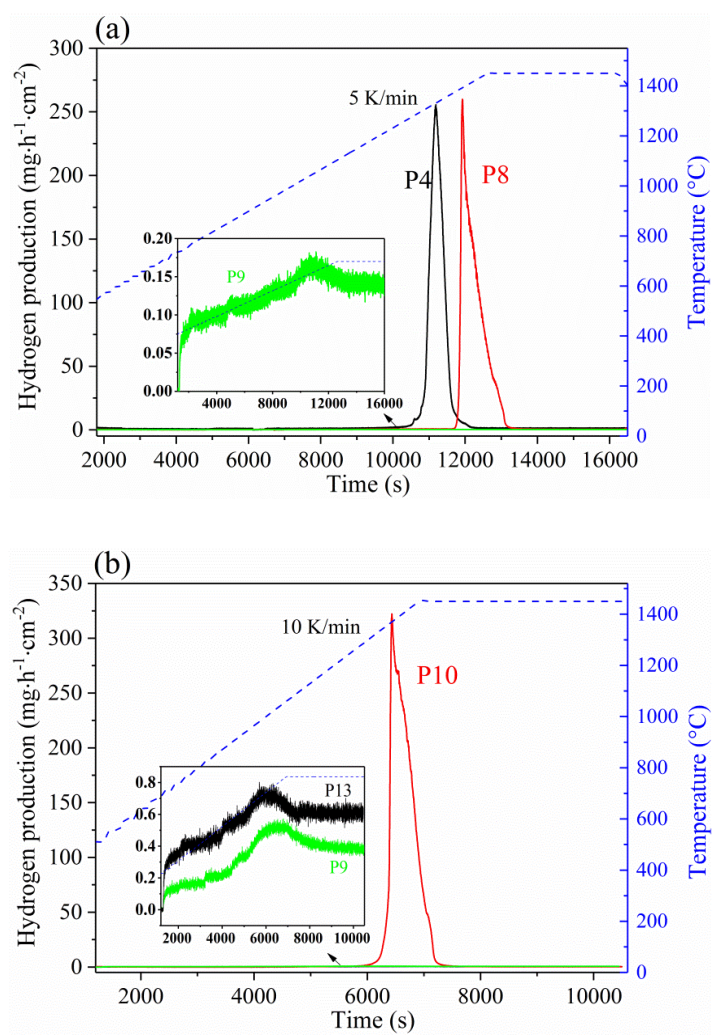

Fig. 2 Typical hydrogen production rates during the transient tests from 500 to 1450°C with subsequent holding for 1 h at 1450°C (a) 5 K/min and (b) 10 K/min heating rates. Appearance of a high concentration peak of hydrogen

was observed for the failed samples and inserted images showed the extremely low hydrogen release rate for the samples with protective oxide scale.

Table 2 Steam compatibility of FeCrAl alloys as function of composition and heating rate in the transient tests. The maximum tolerable temperature in steam of samples that did not fail was marked as 1450°C in italic bold.

| Sample | Designation composition | Tolerant temperature | |
|---|---|---|---|
| | | 5 K/min | 10 K/min |
| P1 | Fe6Cr6Al | 1335°C | 1319°C |
| P2 | Fe8Cr6Al | 1353°C | 1336°C |
| P3 | Fe10Cr5Al | 1327°C | 1298°C |
| P4 | Fe14Cr4Al | 1253°C | 1231°C |
| P5 | Fe16Cr4Al | 1350°C | 1305°C |
| P6 | Fe6Cr8Al | 1368°C | 1345°C |
| P7 | Fe10Cr7Al | 1356°C | 1331°C |
| P8 | Fe12Cr7Al | 1373°C | 1352°C |
| P9 | Fe16Cr6Al | ***1450°C*** | ***1450°C*** |
| P10 | Fe12Cr5Al | 1353°C | 1333°C |
| P11 | Fe12Cr8Al | ***1450°C*** | ***1450°C*** |
| P12 | Fe16Cr8Al | ***1450°C*** | ***1450°C*** |
| P13 | Fe12Cr5Al0.3Y | ***1450°C*** | ***1450°C*** |
| P14 | Fe16Cr8Al0.3Y | ***1450°C*** | ***1450°C*** |
| K-APM | Fe20Cr5.8Al | ***1450°C*** | 1378°C |
| K-D | Fe20Cr4.8Al | ***1450°C*** | 1372°C |

XRD patterns of three representative samples after the transient tests are displayed in Fig.3. The results confirmed that the alloys which suffered catastrophic oxidation (like P8) were completely oxidized and converted to $Fe_3O_4$ and $Fe(Cr,Al)_2O_4$ oxides. The protective scale formed on the samples which survived during the transient tests (like P9 and P13) was $α-Al_2O_3$. In these cases, the diffraction peaks belonging to Fe(Cr) phase were generated by the alloy substrates.

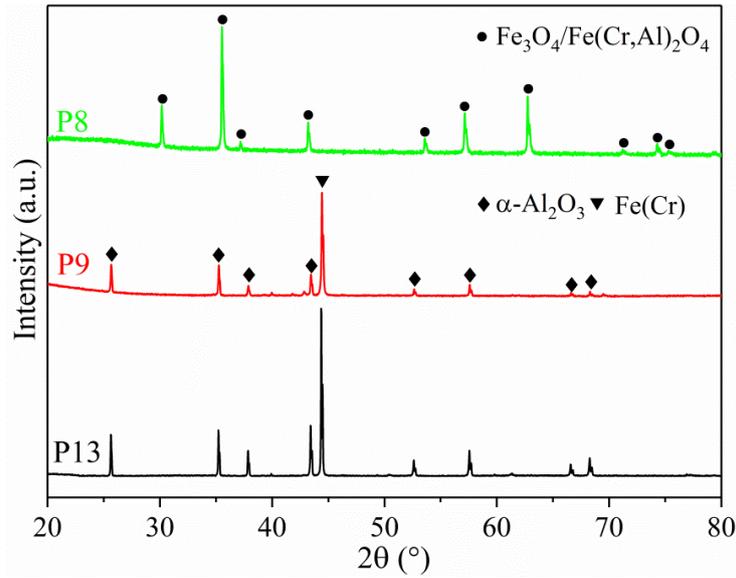

Fig.3 XRD patterns of three representative samples P8-Fe12Cr7Al, P9-Fe16Cr6Al and P13-Fe12Cr5AlY after transient tests with 10 K/min heating rate up to 1450°C.

The surface aspect and cross-sectional morphology of the samples exposed to the transient tests were evaluated using SEM/EDS characterization methods. Fig.4 and Fig.5 present micrographs of some representative samples tested with 10 K/min heating rate. The model alloys that encountered catastrophic oxidation (P1-P8, P10) displayed similar surface structure (example P8 in Fig.4a) with rough-granular morphology containing high volume fraction of pores. The absence of a protective oxide scale on the surface is explicitly evident. The commercial alloys, Kanthal-APM and D, which also failed during the tests with 10 K/min heating rate, apparently revealed non-formation of protective alumina scales. However, they displayed a slightly different surface morphology compared to those of the failed model alloys. The surface of Kanthal alloys was smoother with platelet grains (e.g. Kanthal-APM in Fig.4b). This different aspect could be attributed to the different fabrication processes for the model and the commercial alloys.

The survived alloys, having a divergent surface aspect, were protected by an alumina scale (Fig.4c and d). The oxide scale adherence was low for P9 alloys without Y (Fig.4c). Spallation and delamination of the alumina scale were seen, apparently resulting from thermal stress induced by an abrupt temperature change during cooling phase [30]. The alumina scale was convoluted with a thin layer of equiaxed grains at the gas-metal interface and the typical elongated columnar α-$Al_2O_3$ grains underlying. In case of P13 sample with Y addition, the scale was smooth, relatively dense and highly adherent with an exclusively columnar grain structure on the surface (Fig.4d).

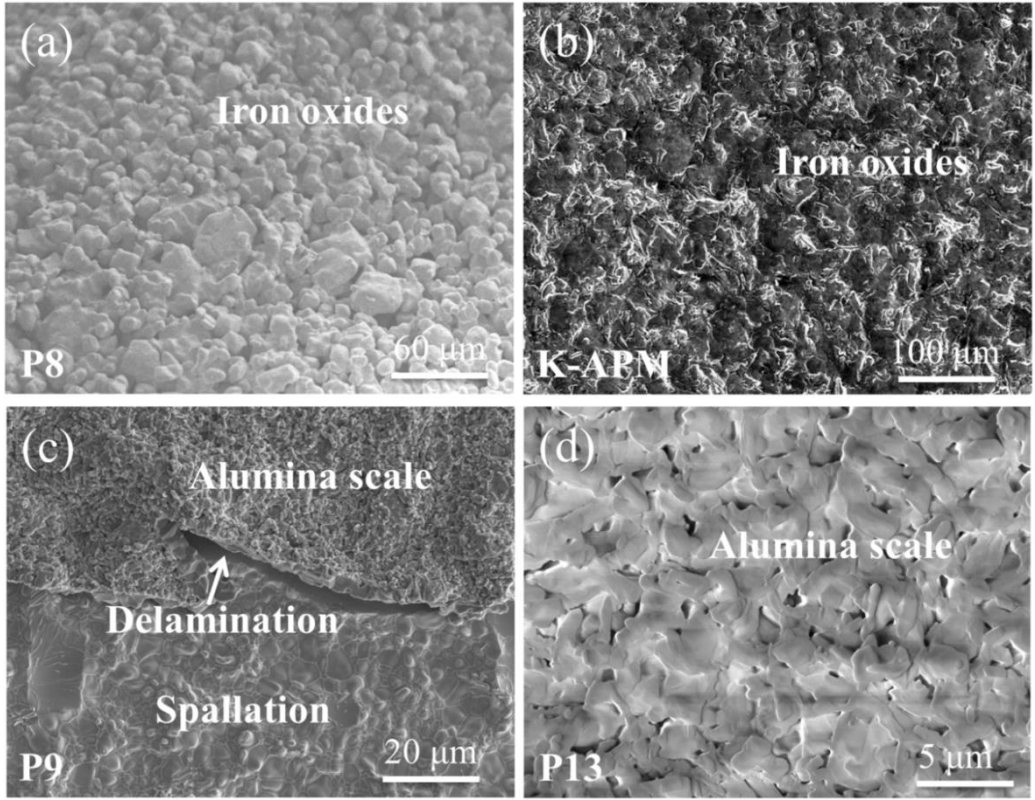

Fig.4 Surface SEM images of representative samples after the transient tests with 10 K/min heating rate up to 1450°C. (a) P8-Fe12Cr7Al, (b) Kanthal-APM, (c) P9-Fe16Cr6Al, (d) P13-Fe12Cr5AlY.

In accordance with previous findings, the cross-sectional SEM images in Fig.5 furthermore confirmed that a protective α-$Al_2O_3$ scale formed on the surface of the survived samples. In the case of sample P9 without Y addition (Fig.5a), the alumina scale showed low adherence with spallation and delamination. Both the ambient/scale interface and the scale/alloy interface became convoluted after oxidation and voids were observed within the scale. The alumina scale grown on the samples with Y addition is highly adherent as can be seen on the micrographs of the samples P13 and P14 (Fig.5b and c). However, slightly convoluted interfaces and some voids were also observed for P13 alloy with relatively low Cr and Al content (Fig.5b). The scale formed on P14 sample was uniform and dense, free of cracks and voids (Fig.5c). Segregation of Y-enriched bright particles on the surface and within the scale was visualized by the contrast discrepancy, as well as ascertained by EDS mapping. The average thickness of the alumina scale on the survived samples was at around 5 μm. A thin region of internal oxidation, adjacent to the oxide scale, was also observed due to the overdoping effect [31].

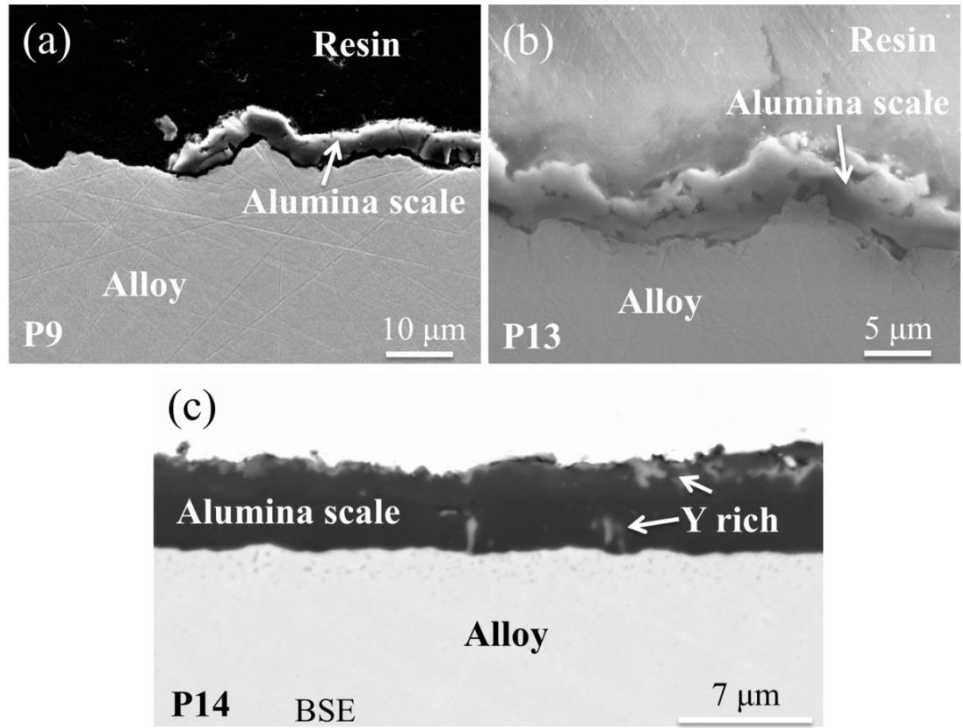

Fig.5 Cross-sectional SEM images of representative samples with protective oxide scale after transient tests with 10 K/min heating rate up to 1450°C. (a) P9-Fe16Cr6Al, (b) P13-Fe12Cr5AlY, (c) P14-Fe16Cr8AlY.

**3.1.2 Experiments up to 1300°C**

In order to gain more insights into the failure mechanism of the samples during the transient tests, tests only up to 1300°C without subsequent holding period were performed on a selection of three samples with various Cr and Al contents: P5, P6 and P10. Their maximum tolerable temperatures in steam determined from previous transient tests slightly exceed 1300°C as shown in Table 2.

As expected, the hydrogen release rate during the tests (not shown here) demonstrated similar tendency as in previous tests up to 1450°C, without abrupt increase below 1300°C for these three alloys. Fig.6 shows the XRD patterns of the samples after the tests with 10 K/min heating rate. The results illustrate that all three samples formed an external α-$Al_2O_3$ scale, as alumina was the principal phase indexed except the substrate. Diffraction peaks with relatively low intensity attributing to the oxide of iron and aluminum ($FeAl_2O_4$) were also observed for P10 sample.

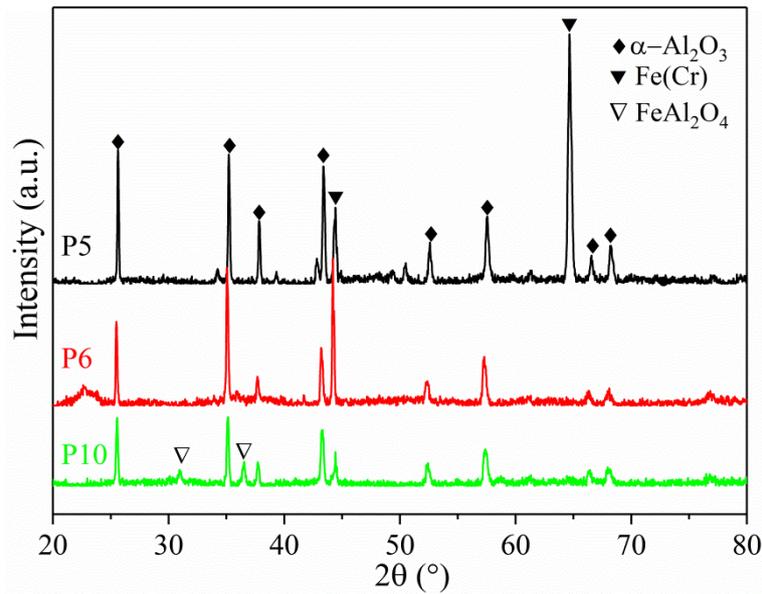

Fig.6 XRD patterns of the three representative samples P5-Fe16Cr4Al, P6-Fe6Cr8Al and P10-Fe12Cr5Al after the transient tests to 1300°C with 10 K/min heating rate.

The surface and cross-sectional SEM micrographs of the samples after the transient tests up to 1300°C are displayed in Fig.7. All the alloys were completely covered with a thin α-$Al_2O_3$ scale except cracking and some spallation of the scales occurred (Fig.7 c and d). All the surfaces were free of large nodular iron-based oxide. Previous studies have demonstrated that FeCrAl alloys containing at least 5 wt.% chromium and 4 wt.% aluminum successfully initiated the growth of oxide scales whereas $Al_2O_3$ acting as the only scale component during short-time high-temperature oxidation [32], and this study was consistent with previous findings.

In general, growth and thermally induced stresses are accumulated within the scale-substrate system during oxidation. Stresses relaxation are frequently triggered through substrate deformation, scale wrinkling and/or spallation [33]. It is necessary to mention that different stress relaxation mechanisms were observed for these three alloys, i.e. P5 and P6 mainly by scale wrinkling and P10 by scale spallation as shown in Fig.7 (upper images). These two features have been proved not to be interrelated and can be significantly influenced by the alloy composition, substrate strength and grain orientation [23]. The scales were composed of fine grains and contained considerable surface porosity. Fine voids (or cavities) were observed within the scales and at the scale/alloy interfaces. Higher void densities were observed in case of the samples tested at higher heating rates, e.g. micrographs of P10 samples tested with 10 K/min (Fig.7c') and with 5 K/min (Fig.7d'). Slightly thicker scales grew on alloys at lower heating rate. For instance, the average thickness of the scales determined from the cross-section images were around 1.35 μm and 1.46 μm for P10 with 10 and 5 K/min heating rate, respectively. However, these $Al_2O_3$ scales are supposed to lose their protective effect with further increasing of temperature, as proved during the transient tests to 1450°C.

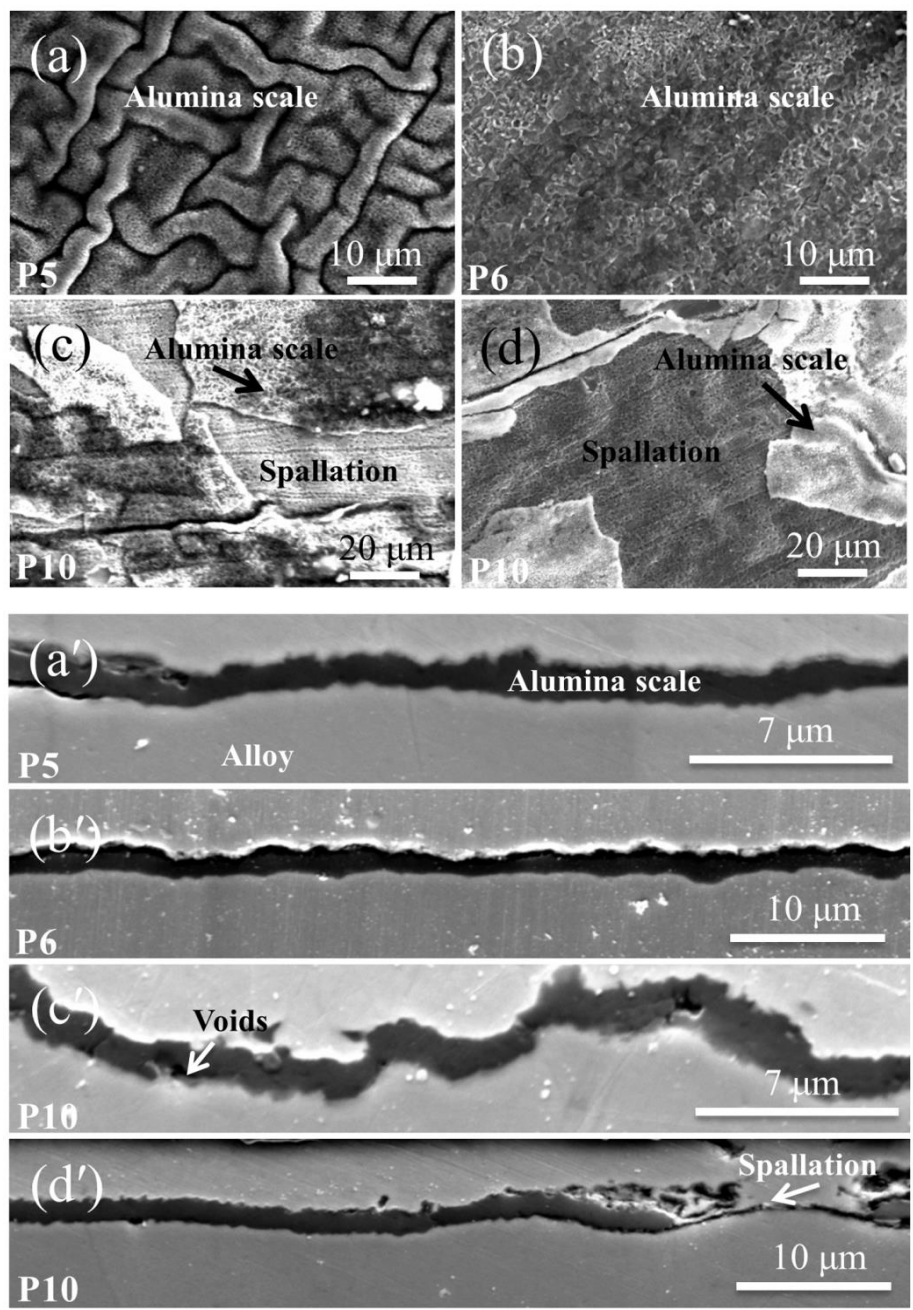

Fig.7 SEM images of surface view (top) and cross-sectional view (bottom) of three representative samples after transient tests to 1300°C. (a) and (a') P5-Fe16Cr4Al, (b) and (b') P6-Fe6Cr8Al, (c) and (c') P10-Fe12Cr5Al with 10 K/min heating rate; (d) and (d') P10 with 5 K/min heating rate.

## 3.2 Isothermal oxidation

Isothermal oxidation tests in the temperature range of 1200°C to 1500°C of P9-P14 model alloys and the two commercial alloys with relatively high Cr and Al content were further performed to simulate extremely fast ramp heating rate scenarios.

Similar to the transient tests, the samples could also be classified into two categories: catastrophic oxidation (marked as "failure") and formation of a protective scale (denoted by "protective") by identifying the hydrogen release rate, post-test appearance as well as the mass gain. Fig.8 shows the typical hydrogen production rate during the isothermal tests at different temperatures for 1 hour with inserted post-test appearance of the model alloy P14 at 1400°C and the commercial alloy Kanthal-APM at 1400°C and 1500°C. The performances of the examined alloys are summarized in Table 3.

Isothermal oxidation test at 1200°C and 1300°C resulted in catastrophic oxidation behavior of the alloy with the lowest Cr and Al content (P10). At higher Cr and Al concentrations, the protective $Al_2O_3$ scale layer was established for P9 and P11. Doping the alloys with reactive element (Y) provided some beneficial effect on improving the tolerance of the alloys to high-temperature steam as P10 alloy failed, whereas P13 with additional 0.3 wt.% Y survived at 1200°C. Nevertheless, both alloys failed at 1300°C.

It is interesting to note that distinctive behaviors appeared once the isothermal temperature reached 1400°C and 1500°C. All the tested samples underwent catastrophic oxidation at 1400°C, but exhibited protective feature at the higher temperature of 1500°C. As displayed in Fig.8a, for all samples that failed at various temperatures, the hydrogen production rates increased sharply to an extremely high level when the steam was introduced into the furnace, then decreased quickly to the level before steam exposure. In case of the alloys, labelled "protective" showing tolerance to high-temperature steam, the hydrogen production rates firstly increased quickly at the initial stage of oxidation and then remained relatively constant or slightly decreased. In addition, the hydrogen flow rates were maintained at a remarkably low level, even at 1500°C, around three orders of magnitudes lower compared to those of the failed samples. The post-test appearance of the model alloy P14 (Fig.8b) and the commercial alloy Kanthal-APM (Fig.8c) after oxidation at 1400°C clearly shows no formation of protective oxide scale on the surface. The samples disintegrated and fractured with loose structure containing grooves and pores, and the color of the surface transferred to deep dark. Both samples displayed significantly high mass gains, around 40 wt.% corresponding to more than 150 mg/cm$^2$. These failure features are in line with the extremely high release of hydrogen during the tests of the two alloys at 1400°C as depicted in Fig.8a. However, the APM alloy after oxidation at 1500°C shows a completely different surface morphology (Fig.8d), i.e. the sample partially melted without maintaining its original geometry (tube) but displaying a compact structure. The recorded mass gain is merely ~1.20 mg/cm$^2$, directly confirming the growth of a protective oxide scale.

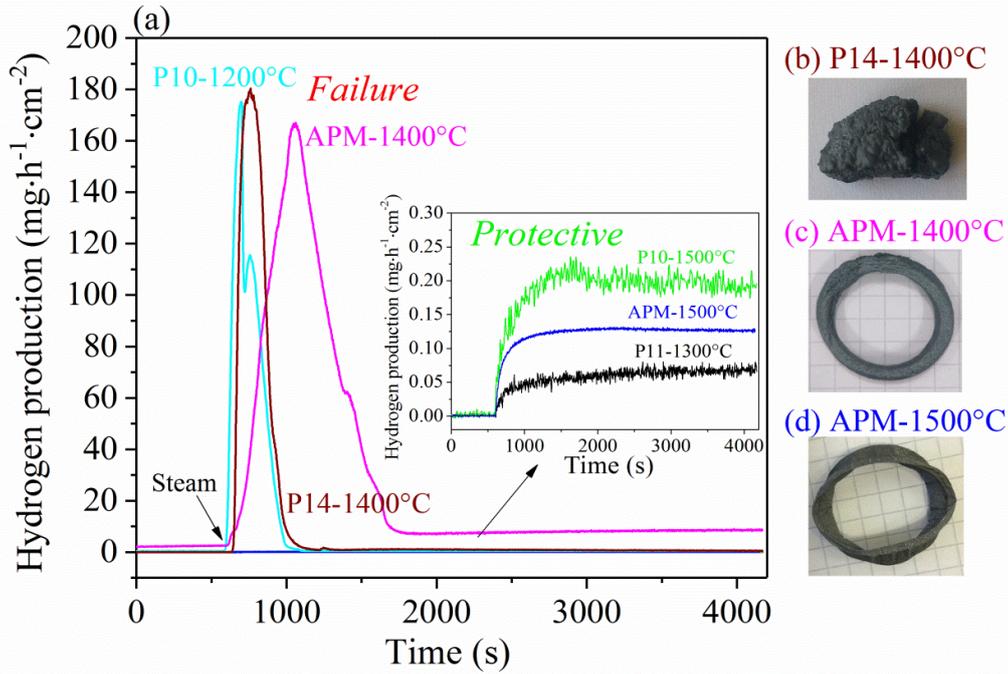

Fig.8 (a) Typical normalized hydrogen production rates during the isothermal tests at different temperatures for 1 h and inserted post-test appearances of (b) model alloy P14 at 1400°C (failure) and commercial alloy Kanthal-APM (c) at 1400°C (failure) and (d) at 1500°C (protective).

Table 3 Compatibility of FeCrAl-based alloys during isothermal test at 1200°C -1500°C in steam for 1 h

| Sample | Designation composition | 1200°C | 1300°C | 1400°C | 1500°C |
|---|---|---|---|---|---|
| P9 | Fe16Cr6Al | - | *Protective* | - | - |
| P10 | Fe12Cr5Al | Failure | Failure | - | *Protective* |
| P11 | Fe12Cr8Al | *Protective* | *Protective* | Failure | - |
| P12 | Fe16Cr8Al | - | - | Failure | *Protective* |
| P13 | Fe12Cr5Al0.3Y | *Protective* | Failure | Failure | - |
| P14 | Fe16Cr8Al0.3Y | - | - | Failure | - |
| K-APM | Fe20Cr5.8Al | - | *Protective* | Failure | *Protective* |
| K-D | Fe20Cr4.8Al | - | *Protective* | Failure | *Protective* |

Fig. 9 shows the typical surface and cross-sectional morphologies of representative FeCrAl alloys after the isothermal oxidation. A protective α-$Al_2O_3$ scale formed for the samples that did not undergo catastrophic oxidation (Fig.9a, b and d). A wrinkled or convoluted appearance of scales with high adherence formed on alloys with Y addition, e.g. P13 at 1200°C (Fig.9a). The alumina scales show low adherence with delamination and spallation for alloys without Y addition (Fig.9b, P11 at 1300°C). At 1400°C, all the tested samples failed to form protective alumina scales and transformed to brittle iron-based oxides with large number of pores (Fig.9c as an example). It is quite surprising that the P10 alloy with relatively low Cr and Al content, which even failed at 1200°C, successfully developed an α-$Al_2O_3$ scale on the surface at 1500°C (Fig.9d). The surface of the sample became rough, like the surface of APM-1500°C in Fig.8d. Even though the scale cracked due to local stresses concentration induced by the change of sample shape, only few spallation was shown. This might be attributed to the stress relaxation easily triggered by plastic deformation of the alloy substrate at such high temperature.

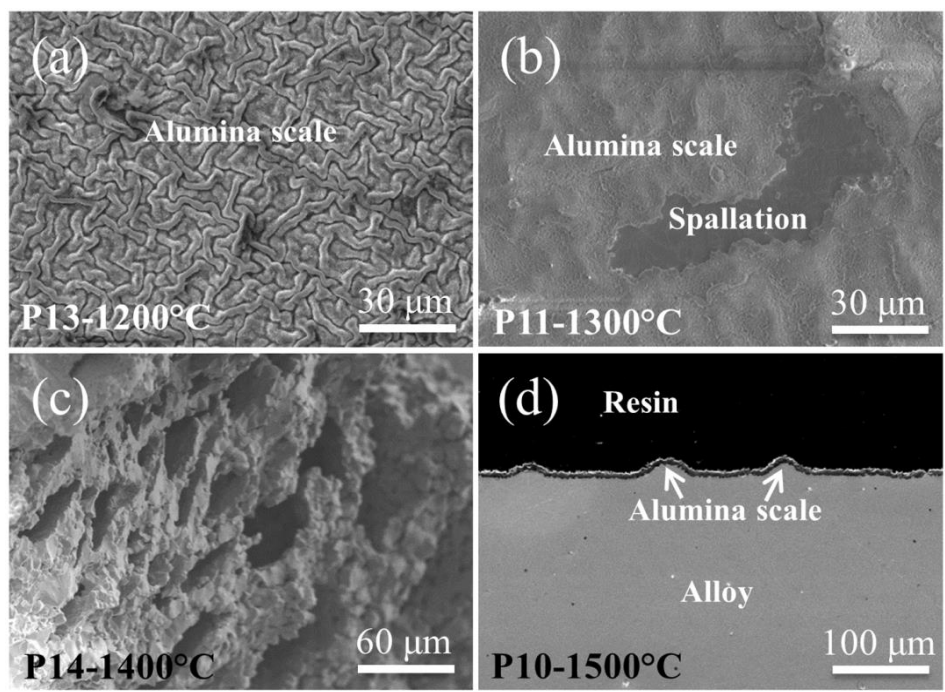

Fig.9 Typical SEM images of surface view (top) and cross-sectional view (bottom) after the isothermal test for 1 h at different temperatures. (a) P13-Fe12Cr5Al0.3Y, (b) P11-Fe12Cr8Al, (c) P14-Fe16Cr8Al0.3Y and (d) P10-Fe12Cr5Al.

Isothermal test at 1400°C and 1500°C in steam for only 15 min were additionally performed for the selected alloys to ascertain the completely dissimilar behaviors at the two temperatures. The same phenomena, namely catastrophic oxidation at 1400°C and protective effect at 1500°C (Fig.8 and Table 3), were further confirmed. Oxidation at 1400°C for 15 min already resulted in full consumption of the alloy substrates. Oxide scales with similar sublayer configuration were developed on alloys tested at 1500°C. Fig.10 shows the structure of the oxide scale grown on P12 (Fe16Cr8Al) alloy and EDS point measurement results after oxidation at 1500°C for 15 min as an example. Four specific areas can be easily distinguished by the discrepancy between their color and morphology as marked in Fig.10a. The outmost layer with the darkest contrast (EDS points 3 and 6) is suggested to be the oxide formed at the very initial oxidation stage that contains the highest content of iron. A porous layer with relatively high Cr content (EDS point 4) grew beneath the surface layer. This layer seems to stem from the transient oxidation stage before the protective alumina scale established [34]. The oxidizing gas penetrating through this layer can be effectively treated as gas transfer in porous media. Selective oxidation of Al prompts the growth of an alumina scale (EDS point 2, 5 and 7) under the transient layer due to alumina remains the most stable oxide in this ternary system. For this yttrium-free alloy, the alumina scale is susceptible to cracking and spallation due to its low adherence (EDS point 1).

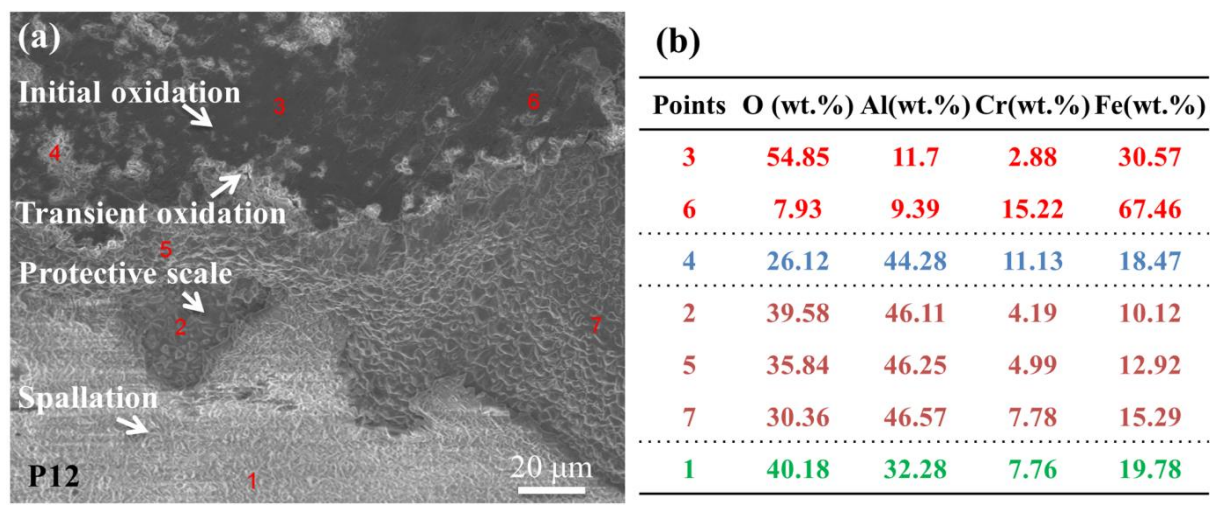

Fig.10 (a) Surface view of the P12-Fe16Cr8Al alloy after isothermal oxidation at 1500°C in steam for 15 min, (b) EDS point measurement results in (a).

## 4. Discussion

Generally, a critical value of Cr and Al addition is required to guarantee the growth of a protective, pure alumina scale on FeCrAl alloys during high-temperature oxidation. Meaningful observations of this study confirmed that the protective effect that can be established or not during the initial stage of oxidation at elevated temperatures is considerably influenced by not only the alloy chemical composition, but also the applied heating schedule.

### 4.1 The influence of Cr and Al content

Oxidation maps are frequently used to illustrate various types of oxide scales formed on the surface of aluminum-containing ternary alloys when exposed to oxidizing conditions. The maps generally contain three regions, corresponding to the alumina-forming, base-element oxide-forming and third-element (usually Cr) oxide-forming oxides [21]. The domain of these three regions fluctuates mainly depending on the oxidation temperature and the chemical environment. The frontier between the alumina- and base-element (here Fe) oxide-forming regions remains of specific interest since it clarifies the critical aluminum concentration needed to form an external protective alumina scale. Based on the above-described results, the domain of protective alumina scale, i.e. the oxide map, has been roughly estimated for the FeCrAl-based alloys exposed to steam during the transient tests performed from 500°C to 1450°C with subsequent holding time of one hour at 1450°C (Fig.11). A $Cr_2O_3$-forming region is substituted by a non-protective Fe-based scale herein, because the $Cr_2O_3$ scale cannot sustain such high temperature.

For yttrium-free alloys, as shown in Fig.11, the dividing threshold curves between the two regions at both different heating rates demonstrate that a synergistic combination of critical Al and Cr concentration is required to prompt the establishment and preserve of protective alumina scales throughout the whole exposure period. The critical concentration of Al declines gradually with increasing Cr content, which indicates the beneficial third element effect (TEE) of Cr on alumina scale growth. The criterion concerning the critical Al concentration ($C_{Al}$) required to form a protective alumina scale on FeCrAl-based alloys without Y addition during the transient tests as a function of heating rate and chromium content ($C_{Cr}$) in the range 6-20 wt.% can be roughly defined as follows:

$$C_{Al} = 13.43 - 0.58(C_{Cr}) + 0.00857(C_{Cr})^2 \text{ [wt.\%], 5 K/min;} \tag{1}$$

$$C_{Al} = 11.76 - 0.32(C_{Cr}) \ (6<C_{Cr}<17) \text{ [wt.\%], 10 K/min,} \tag{2a}$$

$$C_{Al} \approx 6.40 \ (17<C_{Cr}<20) \text{ [wt.\%], 10 K/min.} \tag{2b}$$

These equations correspond to the frontier lines, which separate the protective alumina stability domain from the non-protective Fe-based oxides stability domain in Fig. 11.

Within the investigated chemical composition region (Cr: 6-20 wt.%, Al: 4-8 wt.%), it was observed that Al has a more pronounced beneficial effect on improving the maximum tolerable temperature of FeCrAl alloys in steam than Cr, as demonstrated by the results and displayed in Fig.11. For instance, around 30 K improvement was observed when the Al content increased by 2 wt.%, more than 10 K higher than that of Cr at same increment (by about 18 K improved) when comparing the maximum tolerable temperature of alloys P1-P8 in Table 2. The additional Al content needed ($N_{Al}$) while keeping the Cr content constant for changing the oxidation behavior of alloy P6 from non-protective to protective is around 2 times lower than the increasing Cr content needed for achieving the same effect ($N_{Cr}$), in Fig.11a. Recent investigations, in which Fe-Al and Fe-Cr alloys were compared, also show that in most cases Al is around 1.5 - 2 times more effective as an oxidation retardant than Cr at 1200°C considering scaling losses (weight loss by oxidation) [14]. High Al addition supposes to reduce the density of interfacial voids between the scale and the substrate and to accelerate thickening of the α-$Al_2O_3$ scale [35], resulting in better beneficial outcome than Cr in improving the maximum tolerant temperature during transient tests.

Another meaningful observation of this study is the saturation of the third element effect (TEE) of Cr, especially at higher heating rates, once the Cr content reaches around 20 wt.% (Fig.11). Several different physical mechanisms have been proposed to explain the origin for the observed Cr TEE in FeCrAl alloys. It was suggested that Cr can form protective $Cr_2O_3$ at low temperatures, inhibit the external oxidation of Fe as secondary oxygen getter, influence Al diffusivity and activity in the alloy and accelerate the formation of $Al_2O_3$ by acting as nucleation sites for the α-$Al_2O_3$ [14,35,36]. Recently ab-initio electronic structure calculations demonstrated that Cr acts as an alumina booster. Substituting Fe by Cr in Fe–Al alloys obviously increases the driving force of Al to diffuse from the bulk to the surface. The induced driving force for the diffusion of Al atoms increases substantially when the Cr content in the base alloy is increased from 0 up to 10-15 at.%. Hence, the Al content at the surface simultaneously increases with increasing Cr concentration [14]. The anticipated increased Al-content in the surface region improves the formation as well as increases the volume fraction of alumina during initial oxidation. On the other hand, transient oxidation mainly involves oxidation of Cr near the surface leading to Cr-oxide patches, which can serve as oxidation retardants and nucleation centers for α-$Al_2O_3$ [14,35]. The addition of Cr, as third alloying element that can boost the formation of the Al-oxide scale on the surface, allows a reduction of the Al-content in bulk within an acceptable limit. However, the calculations also demonstrate that the boosting effect of Cr becomes saturated once the Cr concentration reaches around 15 at.%. Additionally, chromium has been proved predominantly to be effective during the initial transient stages of oxidation when the initial nuclei of iron-, chromium-, and aluminum- oxides can grow [32]. In the current study, the transient oxidation tests conducted up to 1300°C (Fig.7) confirmed that all three samples (P5, P6, P10) have been completely covered by a α-$Al_2O_3$ scale. Since only the α-$Al_2O_3$ scale can provide sufficient protection against steam oxidation at temperature above 1200°C, it is reasonable to assume that a saturation of the third element effect (TEE) and a minimum Al content is needed to maintain the protective effect up to 1450°C.

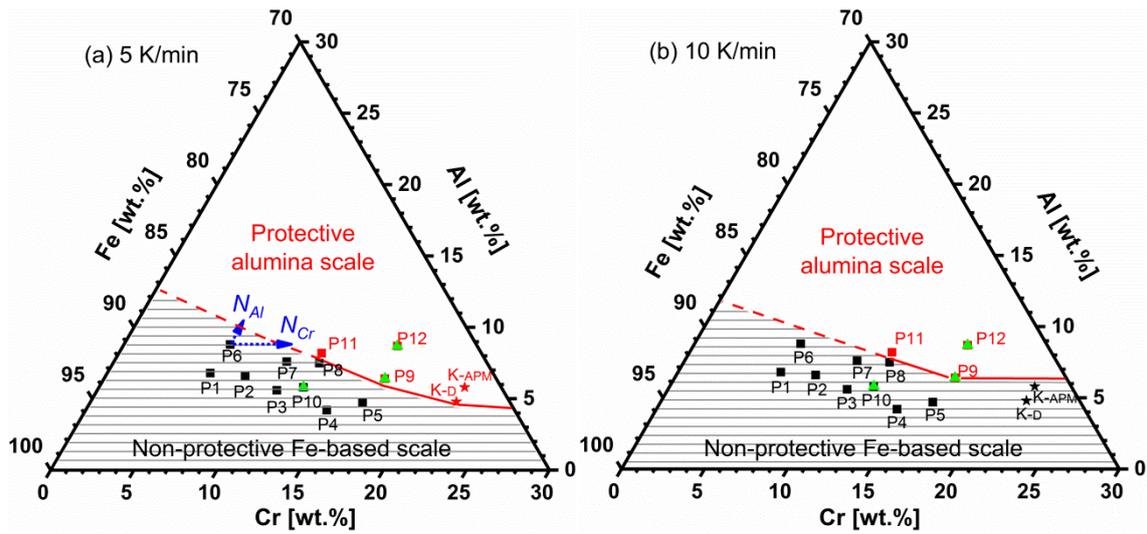

Fig.11 Estimated "oxide map" for the oxidation behavior of FeCrAl-based alloys during the transient tests from 500 to 1450°C with subsequent holding for 1 h at 1450°C in steam. (a) 5 K/min and (b) 10 K/min heating rate during the ramp period. (■) and (★): alloy specimens showing catastrophic oxidation with non-protective Fe-based oxide scale; (■) and (★): alloy specimens forming protective alumina scale; (▲): alloy specimens with Y addition forming protective alumina scale.

**4.2 The influence of heating schedules**

Concerning the oxidation of FeCrAl alloys, numerous studies have confirmed that the oxidation kinetics follow a parabolic or cubic law controlled by diffusional processes in the oxide scale once a protective alumina scale is established [30,32,37–41]. However, transient oxidation processes are often observed during the initial stage of oxidation, and the oxidation kinetics correspondingly diverges from parabolic rate law and tends to obey a linear law [40,42]. Fig.12 summarizes the schematic of the oxide scale architecture and corresponding depth profile of oxidant partial pressure during the initial oxidation stage of FeCrAl alloys at elevated temperatures.

The ramp heat-up period during the transient oxidation test can be viewed as a pre-oxidation process. As shown in Fig.7a and Fig.9a, the alumina oxide scale that grows rapidly and initially, particularly on yttrium-free alloys which were heated rapidly, tends to have a more convoluted and porous structure. Furthermore, metastable alumina and high concentration of base elements, in this case Fe and Cr, are also found to be incorporated into the initially formed alumina scale [41–44]. It is reasonable to assume that the initial alumina scale shows more structural and compositional defects compared to scale growth during the steady-state period. The density and dimension of the cracks and pores in the initially formed alumina scale will also progressively increase because of different thermal expansion coefficients of FeCrAl substrates (~16 × $10^{-6}$/K) and α-$Al_2O_3$ scale (~10 × $10^{-6}$/K) [24] while temperature reaches designated values (Fig.7 c and d). The oxidizing gas penetrating the oxide scale could be transformed from a single diffusion process into a combination of diffusion and gas transfer through porous media with increasing temperature. The latter process (gas transfer) can profoundly enhance the delivery of the gas. Consequently, the partial pressure of the oxidant at the scale/metal interface can remarkably increase (Fig.12). Since the partial pressure of the oxidant at the scale/metal interface is inversely proportional to the scale thickness, the average thickness of the alumina scale additionally appears to be critical for scale susceptibility to failure. Increasing the heating rates during temperature ramp-up period results in growth of a thinner alumina scale owing to the shorter exposure time. The thinner the oxide scale, the higher the oxidant partial pressure at the scale/metal interface at a specific temperature, and vice versa. Once the substrate cannot provide adequate Al to consume the oxidant and the partial pressure at the interface is high enough to oxidize both Fe and Cr as well, the alloys are suggested to suffer from catastrophic oxidation as observed during the transient test. The maximum compatible temperature at which the oxide scales fail, i.e. catastrophic oxidation occurs, declines if a thinner scale grows due to increasing heating rates (Table 2). Recent investigations also reported that FeCrAl(RE) alloys were susceptible to a breakaway oxidation during transient tests at temperature above 1200°C in steam by applying a high ramp rate [27,41]. The occurrence of catastrophic oxidation thus seems to be controlled by a dynamic competition between the solid mass transfer of the Al from the substrate and the transfer of oxidizing gas through the scale toward the metal/scale interface [45,46].

In terms of isothermal oxidation in the temperature range of 1200 - 1400°C that are few tens to hundreds Kelvin lower than the solidus temperature of specific FeCrAl alloys, the atom diffusion in the alloy is governed by solid diffusion through adjacent vacancies or atom transfer via bond breaking [47]. If the sample can supply sufficient Al to establish a continuous and protective alumina scale beneath the transient oxide layer during initial oxidation (Fig.12), then no catastrophic oxidation phenomena will occur. Otherwise, the transient oxidation process is suggested to dominate the oxidation procedure characterized by rapid and complete consumption of the samples, for instance catastrophic oxidation for all tested alloys at 1400°C (Table 3). Due to the absence of a pre-oxidation process, the alloys are more susceptible to catastrophic oxidation in high-temperature steam as demonstrated in Tables 2 and 3. However, as the temperature continuously increases up to 1500°C, approaching or exceeding the solidus temperature of FeCrAl alloys [48], the diffusion in the substrate will, at least partially, change from solid diffusion into liquid diffusion. The atoms (or clusters) in liquid can move more freely due to higher kinetic energy and larger space compared to those in solid. Typically, the diffusion coefficients in liquid are ~$10^{-5}$ cm$^2$/s. While, the values for solid diffusion are several magnitudes lower than liquid diffusion, in the range of $10^{-11}\sim 10^{-8}$ cm$^2$/s [47,49]. The chemical composition gradient due to selective oxidation of Al drives Al atoms diffusing from the alloys interior to the alloy/scale interfaces. Hence, substantially more Al can diffuse outwardly toward the metal/oxide interface at 1500°C.

As previously mentioned, during oxidation the thermally grown oxide is accompanied by the development of stresses within the oxide scale, comprising of thermal and growth stresses [36]. Thermal stresses induced by an abrupt temperature change arise from the difference in the thermal expansion coefficients between the alloy substrate and its oxide. Growth stresses are associated with the strain in the oxide scale that evolves during its growth keeping at constant temperatures [33]. A dynamic interplay exists between stress generation and relaxation over oxidation time. The relaxation mechanisms can be creep of the oxide scale and/or the alloy substrate. It has been proved that the relaxation of the initial tensile stress in the oxide scales formed on FeCrAl alloys is dominated by creep in the α-Al$_2$O$_3$ scale at 1200°C [33]. It is hypothesized that this mechanism facilitates microcracking, wrinkling and spallation of the oxide scale (Fig.9a and b). However, at an extremely high temperature (1500°C), the stress relaxation mainly occurs via pronounced creep and plastic deformation of the alloy substrate, allowing for an accommodation of stresses within the oxide scale characterized by a compact structure and rare spallation as shown in Fig.9d. It is believed that the above two discussed prerequisites (Al diffusion and stress relaxation) play the determining roles for successful growing of a protective alumina scale at 1500°C in the isothermal tests.

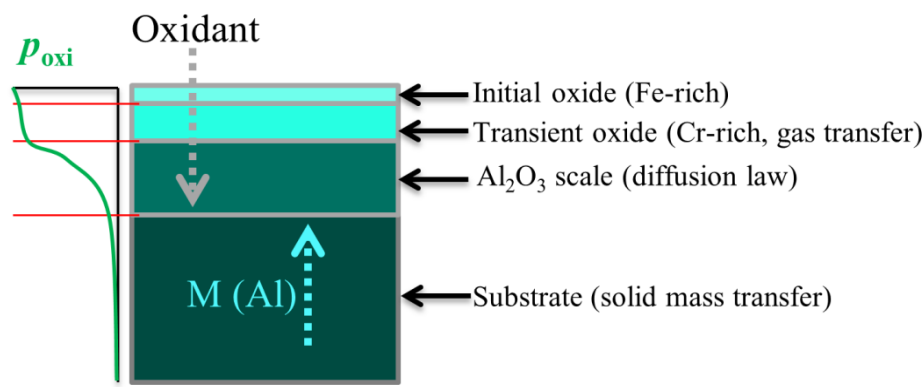

Fig.12 Proposed schematic illustration of oxide scale architecture and corresponding depth profile of oxidant partial pressure during early stage high-temperature oxidation of FeCrAl alloys.

### 4.3 The influence of reactive elements (RE)

The beneficial effects of reactive elements, by means of e.g. changing scale growth mechanism from both anion and cation to predominant anion diffusion, suppressing convoluted morphology and void formation, enhancing selective oxidation and faster scale nucleation as well as remarkably improving the scale adherence, have been widely confirmed and characterized [24]. Similar beneficial effects of REs, particularly during the transient tests, were observed in this study. The oxide scales on yttrium-containing alloys appear smooth, compact as well as highly adherent. With respect to the un-doped alloys, it is typical to see a highly convoluted structure, large interfacial voids and wide-area spallation of the thermally grown oxide scales in case of the alloys survive. Thus, compared to the un-doped alloys with identical base alloy compositions, the maximum tolerable temperature of doped yttrium-containing alloys significantly increased during the transient tests (Table 2). However, only weak improvement was seen during the isothermal tests (Table 3).

It is postulated that REs function effectively through incorporating into the external oxide scale in the form of oxide precipitates primarily at alumina grain boundaries [24,30,50]. The incorporation process is kinetically favored through nucleation and diffusional growth. The morphology and distribution of the precipitates were proved being dependent on the RE mobility in the alloy, RE reservoirs and the oxidation temperature [24,50]. For instance, the incorporation of Hf into the scale was found occurring at a slower rate than that of Zr [50]. With respect to the transient tests herein, it seems that sufficient yttrium is becoming incorporated into the alumina scale and to tremendously improve the scale adherence and prohibit the interfacial void formation as well as reduce the oxidant inward diffusion rate at both heating rates to 1450°C (Fig.5c). However, the isothermal tests do not allow the yttrium to function efficiently due to the lack of a pre-oxidized alumina scale. The limited improvement observed in the isothermal test is probably attributed to the enhanced selective oxidation of Al and faster scale nucleation at the initial oxidation stage due to yttrium addition [24].

It is necessary to mention that the Y concentration in the alloys investigated here is around 0.3 wt%, about 3 times higher than in typical FeCrAlRE alloys (~0.1 wt%). The higher Y content here enables a larger Y reservoir, thereby resulting in accelerating incorporation rate and increasing precipitate density of yttrium oxide. Whereas, the high-level of Y addition also leads to internal oxidation due to over-doped effect. Further research is necessary to optimize the amount of RE dopant and to figure out the relationship between RE reservoir (or concentration) and tolerable temperature of FeCrAlRE alloys under variable test conditions.

## 5. Conclusions

Meaningful observations in this study confirmed that the protective effect arising from the growth of an alumina scale for FeCrAl-based alloys during the initial stage of oxidation at elevated temperatures in steam is not only determined by the concentration of major alloy elements (Al and Cr), reactive element doping, but also considerably influenced by the applied heating schedules. The main conclusions are as follows:

(1) Two distinct phenomena, catastrophic oxidation (rapidly and completely oxidized samples) and protective effect (growth of an external alumina scale) were confirmed during high-temperature steam oxidation of FeCrAl-based alloys in transient tests and isothermal tests.

(2) Increasing the aluminum and chromium contents in the alloys and decreasing the heating rate during ramp period improve the alloys resistance to steam in transient tests; the maximum tolerable temperatures in steam simultaneously rise. Aluminum was found to be more effective for improving the maximum tolerable temperature than chromium.

(3) Isothermal oxidation at 1200°C and 1300°C resulted in catastrophic oxidation of alloys with low Cr and Al content. All the tested samples underwent catastrophic oxidation at 1400°C, however, exhibited protective feature at higher temperature 1500°C.

(4) The reactive element (Y) can significantly improve the high-temperature steam resistance of FeCrAl-based alloys during the transient tests; however, yttrium cannot function efficiently in the isothermal tests probably due to the lack of a pre-oxidized alumina scale.

(5) The occurrence of catastrophic oxidation or not is suggested to be controlled by a dynamic competition between mass transfer of Al from the substrate and transport of oxidizing gas through the scale both toward the metal/oxide scale interface.

**Acknowledgment**

This work was supported by the Helmholtz program NUSAFE at the Karlsruhe Institute of Technology and partially funded by the EC Horizon 2020 project - IL TROVATORE (grant 740415). C. Tang appreciates the PhD fellowship supported by the China Scholarship Council (CSC No.201406080013). The authors also thank U. Stegmaier and P. Severloh for their support during experiments.

# Reference

[1]  A.T. Motta, A. Couet, R.J. Comstock, Corrosion of Zirconium Alloys Used for Nuclear Fuel Cladding, Annu. Rev. Mater. Res. 45 (2015) 311–343. doi:10.1146/annurev-matsci-070214-020951.

[2]  S.J. Zinkle, K.A. Terrani, J.C. Gehin, L.J. Ott, L.L. Snead, Accident tolerant fuels for LWRs: A perspective, J. Nucl. Mater. 448 (2014) 374–379. doi:10.1016/j.jnucmat.2013.12.005.

[3]  K. Barrett, S. Bragg-Sitton, D. Galicki, Advanced LWR Nuclear Fuel Cladding System Development Trade-off Study, 2012.

[4]  M. Steinbrück, N. Vér, M. Große, Oxidation of advanced zirconium cladding alloys in steam at temperatures in the range of 600-1200 °C, Oxid. Met. 76 (2011) 215–232. doi:10.1007/s11085-011-9249-3.

[5]  M. Steinbrück, M. Große, L. Sepold, J. Stuckert, Synopsis and outcome of the QUENCH experimental program, Nucl. Eng. Des. 240 (2010) 1714–1727. doi:10.1016/j.nucengdes.2010.03.021.

[6]  Z. Duan, H. Yang, Y. Satoh, K. Murakami, S. Kano, Z. Zhao, et al., Current status of materials development of nuclear fuel cladding tubes for light water reactors, Nucl. Eng. Des. 316 (2017) 131–150. doi:10.1016/j.nucengdes.2017.02.031.

[7]  K. Terrani, Accident tolerant fuel cladding development: Promise, status, and challenges, J. Nucl. Mater. 501 (2018) 13–30. doi:10.1016/j.jnucmat.2017.12.043.

[8]  C. Tang, M. Stueber, H.J. Seifert, M. Steinbrueck, Protective coatings on zirconium-based alloys as accident-tolerant fuel (ATF) claddings, Corros. Rev. 35 (2017) 141–166. doi:10.1515/corrrev-2017-0010.

[9]  C. Tang, M. Steinbrück, M. Große, T. Bergfeldt, H.J. Seifert, Oxidation behavior of $Ti_2AlC$ in the temperature range of 1400 °C–1600 °C in steam, J. Nucl. Mater. 490 (2017) 130–142. doi:10.1016/j.jnucmat.2017.03.016.

[10]  B. Cheng, Y.J. Kim, P. Chou, Improving Accident Tolerance of Nuclear Fuel with Coated Mo-alloy Cladding, Nucl. Eng. Technol. 48 (2016) 16–25. doi:10.1016/j.net.2015.12.003.

[11]  H.G. Kim, J.H. Yang, W.J. Kim, Y.H. Koo, Development Status of Accident-tolerant Fuel for Light Water Reactors in Korea, Nucl. Eng. Technol. 48 (2016) 1–15. doi:10.1016/j.net.2015.11.011.

[12]  K.A. Terrani, S.J. Zinkle, L.L. Snead, Advanced oxidation-resistant iron-based alloys for LWR fuel cladding, J. Nucl. Mater. 448 (2014) 420–435. doi:10.1016/j.jnucmat.2013.06.041.

[13]  S. Canovic, J. Engkvist, F. Liu, H. Lai, H. Götlind, K. Hellström, et al., Microstructural Investigation of the Initial Oxidation of the FeCrAlRE Alloy Kanthal AF in Dry and Wet $O_2$ at 600 and 800°C, J. Electrochem. Soc. 157 (2010) C223. doi:10.1149/1.3391447.

[14]  E. Airiskallio, E. Nurmi, M.H. Heinonen, I.J. Väyrynen, K. Kokko, M. Ropo, et al., High temperature oxidation of Fe-Al and Fe-Cr-Al alloys: The role of Cr as a chemically active element, Corros. Sci. 52 (2010) 3394–3404. doi:10.1016/j.corsci.2010.06.019.

[15]  Y. Yamamoto, B. Pint, K.A. Terrani, K. Field, S. Maloy, J. Gan, Development and property evaluation of nuclear grade wrought FeCrAl fuel cladding for light water reactors, J. Nucl. Mater. 467 (2015) 703–716. doi:10.1016/j.jnucmat.2015.10.019.

[16]  S.A. Briggs, P.D. Edmondson, K.C. Littrell, Y. Yamamoto, R.H. Howard, C.R. Daily, et al., A combined APT and SANS investigation of a´ phase precipitation in neutron-irradiated model FeCrAl alloys, Acta Mater. 129 (2017) 217–228. doi:10.1016/j.actamat.2017.02.077.

[17]  M.N. Gussev, K.G. Field, Y. Yamamoto, Design, Properties, and Weldability of Advanced Oxidation-Resistant FeCrAl Alloys, Mater. Des. 129 (2017) 227–238. doi:10.1016/j.matdes.2017.05.009.

[18]  B.A. Pint, K.A. Terrani, M.P. Brady, T. Cheng, J.R. Keiser, High temperature oxidation of fuel cladding candidate materials in steam–hydrogen environments, J. Nucl. Mater. 440 (2013) 420–427. doi:10.1016/j.jnucmat.2013.05.047.

[19]  K.G. Field, X. Hu, K.C. Littrell, Y. Yamamoto, L.L. Snead, Radiation tolerance of neutron-irradiated model

Fe–Cr–Al alloys, J. Nucl. Mater. 465 (2015) 746–755. doi:10.1016/j.jnucmat.2015.06.023.

[20] K.A. Terrani, B.A. Pint, Y.J. Kim, K.A. Unocic, Y. Yang, C.M. Silva, et al., Uniform corrosion of FeCrAl alloys in LWR coolant environments, J. Nucl. Mater. 479 (2016) 36–47. doi:10.1016/j.jnucmat.2016.06.047.

[21] Z.G. Zhang, F. Gesmundo, P.Y. Hou, Y. Niu, Criteria for the formation of protective Al2O3 scales on Fe-Al and Fe-Cr-Al alloys, Corros. Sci. 48 (2006) 741–765. doi:10.1016/j.corsci.2005.01.012.

[22] F. Liu, H. Götlind, J.E. Svensson, L.G. Johansson, M. Halvarsson, Early stages of the oxidation of a FeCrAlRE alloy (Kanthal AF) at 900 °C: A detailed microstructural investigation, Corros. Sci. 50 (2008) 2272–2281. doi:10.1016/j.corsci.2008.05.019.

[23] I. Gurrappa, S. Weinbruch, D. Naumenko, W.J.J. Quadakkers, Factors governing breakaway oxidation of FeCrAl-based alloys, Mater. Corros. 51 (2000) 224–235.

[24] D. Naumenko, B.A. Pint, W.J. Quadakkers, Current Thoughts on Reactive Element Effects in Alumina-Forming Systems: In Memory of John, Oxid. Met. 86 (2016) 1–43. doi:10.1007/s11085-016-9625-0.

[25] B.A. Pint, K.A. Terrani, Y. Yamamoto, L.L. Snead, Material Selection for Accident Tolerant Fuel Cladding, Metall. Mater. Trans. E. 2 (2014) 190–196. doi:10.1007/s40553-015-0056-7.

[26] B.A. Pint, Performance of FeCrAl for accident-tolerant fuel cladding in high-temperature steam, Corros. Rev. 35 (2017) 167–175. doi:10.1515/corrrev-2016-0067.

[27] B.A. Pint, K.A. Unocic, K.A. Terrani, Effect of steam on high temperature oxidation behaviour of alumina-forming alloys, Mater. High Temp. 32 (2015) 28–35. doi:10.1179/0960340914Z.00000000058.

[28] C. Tang, M. Steinbrueck, M. Grosse, A. Jianu, A. Weisenburger, H.J. Seifert, High-Temperature oxidation behavior of kanthal APM and D alloys in steam, in: Int. Congr. Adv. Nucl. Power Plants, 2016: pp. 2113–2119.

[29] L.J. Ott, K.R. Robb, D. Wang, Preliminary assessment of accident-tolerant fuels on LWR performance during normal operation and under DB and BDB accident conditions, J. Nucl. Mater. 448 (2014) 520–533. doi:10.1016/j.jnucmat.2013.09.052.

[30] F.A. Golightly, F.H. Stott, G.C. Wood, The influence of yttrium additions on the oxide-scale adhesion to an iron-chromium-aluminum alloy, Oxid. Met. 10 (1976) 163–187. doi:10.1007/BF00612158.

[31] B.A. Pint, Optimization of Reactive-Element Additions to Improve Oxidation Performance of Alumina-Forming Alloys, J. Am. Ceram. Soc. 86 (2003) 686–95. doi:10.1111/j.1151-2916.2003.tb03358.x.

[32] P. Tomaszewicz, G.R. Wallwork, The oxidation of high-purity iron-chromium-aluminum alloys at 800°C, Oxid. Met. 20 (1983) 75–109. doi:10.1007/BF00662042.

[33] P.F. Tortorelli, E.D. Specht, K.L. More, P.Y. Hou, Oxide growth stress measurements and relaxation mechanisms for alumina scales grown on FeCrAlY, Mater. Corros. 63 (2012) 857–861. doi:10.1002/maco.201206760.

[34] S. Yoneda, S. Hayashi, I. Saeki, S. Ukai, Investigation of Initial Transient Oxidation of Fe–xCr– 6at.%Al Alloys Using Synchrotron Radiation During Heating to 1000 °C in Air, Oxid. Met. 86 (2016) 357–370. doi:10.1007/s11085-016-9641-0.

[35] S. Yoneda, S. Hayashi, S. Ukai, The Transition from Transient Oxide to Protective Al2O3 Scale on Fe–Cr–Al Alloys During Heating to 1000 °C, Oxid. Met. 89 (2018) 81–97. doi:10.1007/s11085-017-9778-5.

[36] F.H. Stott, G.C. Wood, J. Stringer, The influence of alloying elements on the development and maintenance of protective scales, Oxid. Met. 44 (1995) 113–145. doi:10.1007/BF01046725.

[37] S.E. Sadique, A.H. Mollah, M.S. Islam, M.M. Ali, M.H.H. Megat, S. Basri, High-temperature oxidation behavior of iron-chromium-aluminum alloys, Oxid. Met. 54 (2000) 385–400. doi:10.1023/A:1004682316408.

[38] S. Dryepondt, J. Turan, D. Leonard, B.A. Pint, Long-Term Oxidation Testing and Lifetime Modeling of Cast and ODS FeCrAl Alloys, Oxid. Met. 87 (2016) 215–248. doi:10.1007/s11085-016-9668-2.

[39] K. Hellström, N. Israelsson, M. Halvarsson, S. Canovic, J.-E. Svensson, L.-G. Johansson, The Oxide Scales

Formed on a Dispersion-Strengthened Powder Metallurgical FeCrAl Alloy at 900 °C in O2 and in O2 + H2O, Oxid. Met. (2015) 1–19. doi:10.1007/s11085-015-9538-3.

[40] C. Issartel, H. Buscail, S. Chevalier, J. Favergeon, Effect of Yttrium as Alloying Element on a Model Alumina-Forming Alloy Oxidation at 1100 °C, Oxid. Met. (2017) 1–12. doi:10.1007/s11085-017-9750-4.

[41] K.A. Unocic, Y. Yamamoto, B.A. Pint, Effect of Al and Cr Content on Air and Steam Oxidation of FeCrAl Alloys and Commercial APMT Alloy, Oxid. Met. 87 (2017) 431–441. doi:10.1007/s11085-017-9745-1.

[42] H. Götlind, F. Liu, J.-E. Svensson, M. Halvarsson, L.-G. Johansson, The effect of water vapor on the initial stages of oxidation of the FeCrAl alloy Kanthal AF at 900°C, Oxid. Met. 67 (2007) 251–266. doi:10.1007/s11085-007-9055-0.

[43] J.Y. Lee, H.G. Kim, M.R. Choi, C.W. Lee, M.H. Park, K.H. Kim, et al., Microstructural evaluation of oxide layers formed on Fe-22Cr-6Al metallic foam by pre-oxidization, Appl. Surf. Sci. 293 (2014) 255–258. doi:10.1016/j.apsusc.2013.12.144.

[44] K. Hellström, N. Israelsson, N. Mortazavi, S. Canovic, M. Halvarsson, J.-E. Svensson, et al., Oxidation of a Dispersion-Strengthened Powder Metallurgical FeCrAl Alloy in the Presence of O2 at 1,100 °C: The Influence of Water Vapour, Oxid. Met. 83 (2015) 533–558. doi:10.1007/s11085-015-9534-7.

[45] P.Y. Hou, Impurity Effects on Alumina Scale Growth, J. Am. Ceram. Soc. 86 (2003) 660–68. doi:10.1111/j.1151-2916.2003.tb03355.x.

[46] B. Lesage, L. Maréchal, A.M. Huntz, R. Molins, Aluminium Depletion in FeCrAl Alloys During Oxidation, Defect Diffus. Forum. 194–199 (2001) 1707–1712. doi:10.4028/www.scientific.net/DDF.194-199.1707.

[47] E.L. Cussler, Diffusion : mass transfer in fluid systems, Cambridge University Press, 2009.

[48] J.W. McMurray, R. Hu, S.V. Ushakov, D. Shin, B.A. Pint, K.A. Terrani, et al., Solid-liquid phase equilibria of Fe-Cr-Al alloys and spinels, J. Nucl. Mater. 492 (2017) 128–133. doi:10.1016/j.jnucmat.2017.05.016.

[49] G. Neumann, C. Tuijn, Self-diffusion and Impurity Diffusion in Pure Metals: Handbook of Experimental Data, Elsevier, 2011.

[50] D. Naumenko, V. Kochubey, L. Niewolak, A. Dymiati, J. Mayer, L. Singheiser, et al., Modification of alumina scale formation on FeCrAlY alloys by minor additions of group IVa elements, J. Mater. Sci. 43 (2008) 4550–4560. doi:10.1007/s10853-008-2639-5.